\documentclass[preprint,showpacs,preprintnumbers,amsmath,amssymb]{revtex4}%
\usepackage{graphicx}
\usepackage{dcolumn}
\usepackage{bm}
\usepackage{amsmath}
\usepackage{amsfonts}
\usepackage{amssymb}%
\providecommand{\U}[1]{\protect\rule{.1in}{.1in}}
\newcommand{\be}{\begin{equation}}
\newcommand{\en}{\end{equation}}
\newcommand{\bea}{\begin{eqnarray}}
\newcommand{\ena}{\end{eqnarray}}
\begin{document}
\title{Warm $\frac{\lambda}{4}\phi^{4}$ inflationary universe model in light of Planck 2015 results}
\author{Grigorios Panotopoulos}
\email{gpanotop@ing.uchile.cl}
\affiliation{Departamento de F\'{\i}sica, FCFM, Universidad de Chile, Blanco Encalada 2008, Santiago, Chile}
\author{Nelson Videla}
\email{nelson.videla@ing.uchile.cl}
\affiliation{Departamento de F\'{\i}sica, FCFM, Universidad de Chile, Blanco Encalada 2008, Santiago, Chile}
\date{\today}

\begin{abstract}

In the present work we show that warm chaotic inflation characterized by a simple
$\frac{\lambda}{4}\phi^{4}$
self-interaction potential for the inflaton, excluded by current
data in standard cold inflation, and by an inflaton decay rate proportional
to the temperature, is in agreement with the latest Planck data.
The parameters of the model are constrained, and our results show that the
model predicts a negligible tensor-to-scalar ratio in the strong dissipative
regime, while in the weak dissipative regime the tensor-to-scalar ratio can be
large enough to be observed.

\end{abstract}

\maketitle



\section{Introduction}

The inflationary universe has become one of the central paradigms in modern cosmology. This is due to the fact that
many long-standing problems of the Big Bang model, such as the horizon, flatness, homogeneity and monopole problems, find a natural
explanation in the framework of the inflationary universe \cite{R1,R106,R103,R104,R105,Linde:1983gd}. However, the essential
feature of inflation is that it generates a
mechanism to explain the Large-Scale Structure (LSS) of the universe \cite{R2,R202,R203,R204,R205}
and provides a causal interpretation of the origin of the anisotropies observed in the
Cosmic Microwave Background (CMB) radiation\cite{astro,astro2,astro202,Hinshaw:2012aka,Ade:2013zuv,Ade:2013uln}, since
primordial density perturbations may be produced from quantum fluctuations during the inflationary era.

The original ``old inflation'' scenario assumed the inflaton was trapped in a
metastable false vacuum and had to exit to the true vacuum
via a first-order transition \cite{R1,R106}. However, the exit could occur neither gracefully nor completely.
The revised version of inflation was proposed by A. Linde \cite{R103,R104}, and A. Albrecht
and J. Steinhardt \cite{R105} in 1982 referred as ``new inflation''. However, these scenarios suffer
from theoretical problems about the duration of inflation and initial conditions. In 1983,
A. Linde considered the case that the initial conditions for scalar field driving inflation
may be chaotic, which is called ``chaotic inflation'' \cite{Linde:1983gd}. This inflation model can solve the
remaining problems, where the potential was chosen to
 be cuadratic or quartic form, i.e. $\frac{m^2}{2}\phi^2$ or $\frac{\lambda}{4}\phi^4$, terms that are always present in the
 scalar potential of the Higgs sector in all renormalizable gauge field
 theories \cite{pich}
 in which the gauge symmetry is spontaneously broken via the Englert-Brout-Higgs mechanism \cite{higgs}. Such models are interesting for their simplicity, and
 has become one of the most favored, because they
predict a significant amount of tensor perturbations due to the inflaton field gets across the trans-Planckian distance during inflation \cite{Lyth:1996im}.
After that, many kinds of inflationary scenarios have been proposed,
related to supersymmetry (SUSY) theory, brane world, string theory, etc. (for
review, see \cite {Lyth:1998xn,Riotto:2002yw,Bassett:2005xm,Baumann:2014nda} ).

On the other hand, with respect to the dynamical mechanisms of
inflation, the warm inflation scenario, as opposed to the standard
cold inflation, has the attractive feature that it avoids the
reheating period at the end of the accelerated expansion \cite{warm}. During the evolution of warm
inflation dissipative effects are important, and radiation
production takes place at the same time as the  expansion of the
universe. The dissipative effects arise from a friction term
which accounts for the processes of the scalar field dissipating into
a thermal bath. In further relation to these dissipative effects,
the dissipative coefficient $\Gamma$ is a fundamental quantity, which
has been computed from first principles in the context of supersymmetry. In particular, in Ref.\cite{26}, a supersymmetric model
containing three superfields $\Phi$, $X$, and $Y$ has been studied, with a superpotential
$W=\frac{g}{\sqrt{2}}\Phi X^2-\frac{h}{\sqrt{2}}XY^2$, where the scalar components of
the superfields are $\phi$, $\chi$, and $y$ respectively. For a scalar field
with multiplets of heavy and light fields, and different decay mechanisms, it is possible to obtain several
expressions for the  dissipative coefficient $\Gamma$, see e.g.,
\cite{26,28,2802,Zhang:2009ge,BasteroGil:2011xd,BasteroGil:2012cm}.

Following Refs.\cite{Zhang:2009ge,BasteroGil:2011xd},
a general parametrization of the dissipative coefficient
$\Gamma(T,\phi)$ can be written as
\begin{equation}
\Gamma(T,\phi)=a\,\frac{T^{m}}{\phi^{m-1}}, \label{G}%
\end{equation}
where  the parameter $a$ is related   with  the dissipative
microscopic dynamics and the exponent $m$ is an integer. This expression for
 the dissipative coefficient includes different cases studied
 in the literature, depending of the
values of $m$ (see Refs.
\cite{Zhang:2009ge,BasteroGil:2011xd}). Specifically,
for the value $m=3$, i.e., $\Gamma \propto T^3/\phi^2$, the
parameter $a$ corresponds to $0.02h^2{\mathcal{N}}_Y$,
where a generic supersymmetric model with
chiral superfields $\Phi$, $X$, and $Y_i,\,i=1,...{\mathcal{N}}_Y$ has
been considered. This case corresponds to a low temperature
regime, when the mass of the catalyst field $m_{\chi}$ is larger than the
temperature $T$ \cite{BasteroGil:2012cm}. On the other hand, $m=1$, i.e., $\Gamma \propto T$
corresponds to a high temperature regime, where the thermal corrections to the catalyst field mass start to be important,
where $a=0.97 g^2/h^2$ \cite{26}. For $m=0$, the dissipative coefficient represents an
exponentially decaying propagator in the high temperature regime. Finally,
For $m=-1$, i.e., $\Gamma\propto \phi^2/T$, agrees with the non-SUSY case \cite{28,PRD}. Additionally, thermal fluctuations during
the inflationary scenario may play a fundamental  role in
producing the primordial fluctuations \cite{62526,1126}. During the warm
inflationary scenario the density perturbations arise from
thermal fluctuations of the inflaton  and dominate over the
quantum ones. In this form,
an essential  condition for warm inflation to occur is the
existence of a radiation component with temperature $T>H$, since the thermal and quantum
fluctuations are proportional to $T$ and $H$,
respectively\cite{warm,62526,1126}. When the universe heats
up and becomes radiation dominated, inflation ends and the
universe smoothly enters in the radiation
Big-Bang phase\cite{warm}. For a comprehensive review of warm
inflation, see Ref. \cite{Berera:2008ar}.

Upon comparison to the current cosmological and astronomical observations, specially those related with the CMB
temperature anisotropies, it is possible to constrain the inflationary models. In particular, the constraints in the $n_s-r$ plane
give us the predictions of a number of representative inflationary
potentials. Recently, the Planck
collaboration has published new data of enhanced precision of
the CMB anisotropies \cite{Ade:2015lrj} . Here, the Planck full mission
data has improved the upper bound on the tensor-to-scalar ratio
$r_{0.002} < 0.11$($95\%$ CL) which is similar to obtained from  \cite{Ade:2013uln} , in which
$r < 0.12$ ($95\%$ CL). In particular, the $\frac{\lambda}{4}\phi^{4}$ model, which predicts a large value of the tensor-to-scalar
ratio $r$, lies well outside of the joint
$99.7\%$ CL region in the $n_s-r$, so it is ruled out by the data. This result confirms previous
findings from e.g., Hinshaw et al. \cite{Hinshaw:2012aka} in which this
model is well outside the $95\%$ CL for the WMAP 9-year data
and is further excluded by CMB data at smaller scales.

In this way, the goal of the present work is to study the possibility that the $\frac{\lambda}{4}\phi^{4}$ model
can be rescued in the warm inflation scenario and be able to agree with the latest observational data. In order to
achieve this, we consider an inflaton decay rate $\Gamma$ proportional to the temperature, which has been computed in the
context of a high temperature supersymmetric model \cite{26}. We stress that, in previous works (see Ref.\cite{BasteroGil:2012cm}), the authors have also studied the quartic potential in the framework of warm inflation. However, our work is different in two ways. First, contrary to the standard cold inflation where the dynamics is determined only by the inflaton potential, in warm inflation also the dissipative coefficient plays an important role, and here we have considered an expression for it not studied in the previous works. Furthermore, in none of these papers the authors used the contour plots in the $r$ and $n_s$ plane to constrain the parameters of the model they studied. On the contrary, in our work here we have used the latest data from Planck, not available at that time, to put bounds on the parameters of the model we have considered.

The outline of the paper is as follows: The next section
presents a short review of the basics of warm inflation scenario. In Sect. \ref{warmq} we study
the dynamics of warm inflation for our quartic potential,
in the weak and strong dissipative regimes; specifically,
we obtain analytical expressions for the slow-roll
parameters and the dissipative coefficient.
Immediately, we compute the cosmological perturbations in
both dissipative regimes, obtaining expressions for the
inflationary observables such as the scalar power spectrum, the
scalar spectral index, and the tensor-to-scalar ratio. Finally,
Sect.\ref{conclu} summarizes our results and exhibits our conclusions.
We choose units so that $c=\hbar=1$.

\section{Basics of warm inflation scenario}\label{warmrew}

\subsection{Background evolution}

We start by considering a spatially flat Friedmann-Robertson-Walker (FRW) universe
containing a self-interacting inflaton scalar field $\phi$ with energy density and pressure
given by $\rho_{\phi}=\dot{\phi}^2/2+V(\phi)$ and $P_{\phi}=\dot{\phi}^2/2-V(\phi)$, respectively,
and a radiation field with energy density $\rho_{\gamma}$. The corresponding Friedmann equations reads
\begin{equation}
H^2=\frac{1}{3M^2_p}(\rho_{\phi}+\rho_{\gamma}),\label{Freq}
\end{equation}
where $M_p=\frac{1}{\sqrt{8\pi G}}$ is the reduced Planck mass.

The dynamics of $\rho_{\phi}$ and $\rho_{\gamma}$ is described by the equations \cite{warm}
\begin{equation}
\dot{\rho_{\phi}}+3\,H\,(\rho_{\phi}+P_{\phi})=-\Gamma \dot{\phi}^{2},
\label{key_01}%
\end{equation}
and
\begin{equation}
\dot{\rho}_{\gamma}+4H\rho_{\gamma}=\Gamma \dot{\phi}^{2}, \label{key_02}%
\end{equation}
where the  dissipative coefficient $\Gamma>0$ produces the decay of the scalar
field into radiation. Recall that this
decay rate can be assumed  to be a function of the
temperature of the thermal bath $\Gamma(T)$, or a function of the
scalar field $\Gamma(\phi)$, or a function of $\Gamma(T,\phi)$ or
simply a constant\cite{warm}.

During warm inflation, the energy density related to
the scalar field predominates  over the energy density of the
radiation field, i.e.,
$\rho_\phi\gg\rho_\gamma$\cite{warm,62526,6252602,6252603,6252604}, but even if small when compared to the inflaton energy density
it can be larger than the expansion rate with $\rho_{\gamma}^{1/4}>H$. Assuming thermalization, this translates roughly
into $T>H$, which is the condition for warm inflation to occur.

When $H$, $\phi$, and $\Gamma$ are slowly varying, which is a good
approximation during inflation, the production of radiation becomes quasi-stable, i.e., $\dot{\rho
}_{\gamma}\ll4H\rho_{\gamma}$ and $\dot{\rho}_{\gamma}\ll\Gamma\dot{\phi}^{2}%
$, see Refs.\cite{warm,62526,6252602,6252603,6252604}. Then, the equations of motion reduce to
\begin{equation}
3\,H\,(1+R)\dot{\phi}\simeq -V_{,\phi},
\label{key_01n}%
\end{equation}
where $,\phi$ denotes differentiation with respect to inflaton, and
\begin{equation}
4H\rho_{\gamma}\simeq \Gamma\,\dot{\phi}^{2}, \label{key_02n}%
\end{equation}
where $R$ is the dissipative ratio defined as
\begin{equation}
R\equiv\frac{\Gamma}{3H}.
\end{equation}

In warm inflation, we can distinguish between two possible scenarios, namely the weak and strong dissipative regimes, defined as $R\ll 1$ and $R\gg 1$, respectively. In the weak dissipative regime, the Hubble damping is still the dominant term, however, in the strong dissipative regime, the dissipative coefficient $\Gamma$ controls the damped evolution of the inflaton field.

If we consider thermalization, then the energy density of the radiation field could be written as $\rho_{\gamma}=C_{\gamma}\,T^{4}$, where the constant  $C_{\gamma}=\pi^{2}\,g_{\ast}/30$. Here,   $g_{\ast}$ represents the number
of relativistic degrees of freedom. In the Minimal Supersymmetric Standard Model (MSSM), $g∗ = 228.75$ and $C_{\gamma} \simeq 70$ \cite{62526}. Combining Eqs.(\ref{key_01n}) and (\ref{key_02n}) with $\rho_{\gamma}\propto\,T^{4}$, the temperature of the
thermal bath becomes
\begin{equation}
T=\left[\frac{\Gamma\,V_{,\phi}^2}{36 C_{\gamma}H^3(1+R)^2}\right]^{1/4}.\label{temp}
\end{equation}

 On the other hand, the consistency conditions for the approximations to hold imply that a set of slow-roll conditions must be satisfied for a prolonged period of inflation to take place. For warm inflation, the slow-roll parameters are \cite{26,62526}
\begin{equation}
\epsilon=\frac{M^2_p}{2}\left(\frac{V_{,\phi}}{V}\right)^2,\,\,\,\eta=M^2_p\left(\frac{V_{,\phi \phi}}{V}\right),\,\,\,\beta=M^2_p \left(\frac{\Gamma_{,\phi}\,V_{,\phi}}{\Gamma\,V}\right),\,\,\,\sigma = M_p^2 \left(\frac{V_{,\phi}}{\phi V}\right).\label{srparam}
\end{equation}

The slow-roll conditions for warm inflation can be expressed as \cite{26,62526}
\begin{equation}
\epsilon \ll 1+R,\,\,\,\eta \ll 1+R,\,\,\,\beta \ll 1+R,\,\,\,\sigma\ll 1+R\label{srcon}
\end{equation}

When one these conditions is not longer satisfied, either the motion of the inflaton is no
longer overdamped and slow-roll ends, or the radiation becomes comparable to the inflaton energy density. In this way,
inflation ends when one of these parameters become the order of $1+R$.

From first principles in quantum field theory, the dissipative coefficient
$\Gamma$ has been computed. As we have seen in the introduction, the parametrization given by Eq.(\ref{G})
includes different cases, depending of the
values of $m$. Concretely , for $m=3$, for which $\Gamma= a T^3\phi^{-2}$, the parameter
$a$ agrees with  $a=0.02\,h^{2}\,\mathcal{N}_Y$,
where a generic supersymmetric model with chiral superfields
$\Phi$, $X$ and $Y_i$, $i=1,...\mathcal{N}_Y$ has been considered. In particular, this inflation ratio decay has been
studied extensively in the literature \cite{BasteroGil:2012cm,Herrera:2015aja}, including the quartic potential \cite{Berera:2008ar}
. For the special case $m=1$,
the dissipative coefficient $\Gamma\propto T$ is related with the high temperature supersymmetry
(SUSY) case \cite{26}. Finally, for the cases $m=0$ and $m=-1$, $\Gamma$
 represents an exponentially decaying propagator in
the high temperature SUSY model and
 the non-SUSY case, respectively\cite{28,PRD}.

\subsection{Perturbations}

In the warm inflation scenario, a
thermalized radiation component is present with $T>H$, then the inflaton fluctuations
$\delta \phi$ are predominantly thermal instead quantum. In this way, following \cite{62526,1126,Berera:2008ar}, the
amplitude of the power spectrum of the curvature perturbation is given by

\begin{equation}
{\cal{P}_{\cal{R}}}^{1/2}\simeq \left(\frac{H}{2\pi}\right) \left(\frac{3H^2}{V_{\phi}}\right)\left(1+R\right)^{5/4}\left(\frac{T}{H}\right)^{1/2},\label{PR}
\end{equation}
where the normalization has been chosen in order to recover the standard cold inflation result when $R\rightarrow 0$ and $T \simeq H$.

By the other hand, the scalar spectral index $n_s$ is given by \cite{62526}
\begin{equation}
n_s=1+\frac{d{\cal{P}_{\cal{R}}}}{d\ln k}\simeq 1+\frac{1}{1+R}\left[-(2-5A_{R})\epsilon-3A_R\eta+(2+4A_R)\sigma\right],\label{nsw}
\end{equation}
where $A_R=\frac{R}{1+7R}$.

Regarding to tensor perturbations, these do not couple to the thermal background, so gravitational waves are only generated by quantum fluctuations, as
in standard inflation \cite{Taylor:2000ze}. However, the tensor-to-scalar ratio $r$ is modified with respect to standard cold inflation, yielding \cite{Berera:2008ar}
\begin{equation}
r\simeq \left(\frac{H}{T}\right)\frac{16\epsilon}{(1+R)^{5/2}}.\label{rwi}
\end{equation}
We can see that warm inflation predicts a tensor-to-scalar ratio suppressed by a factor $(T/H)(1 + R)^{
5/2} > 1$ compared with standard cold inflation.

When a specific form of the scalar potential and the dissipative coefficient $\Gamma$ are considered, it is possible to study the
background evolution under the slow-roll regime and the primordial perturbations in order to
test the viability of warm inflation. In the following we will study how an inflaton decay rate proportional
to the temperature, corresponding to the case $m=1$, influences the inflationary dynamics for the quartic potential. We
will restrict ourselves to the weak and  strong dissipation
regimes.

\section{Dynamics of warm $\frac{\lambda}{4}\phi^4$ inflation}\label{warmq}

Although inflation is widely accepted as the standard paradigm for the early
universe, it is not a theory yet
as we don't know how to answer the question that naturally arises, "what is
the inflaton and what is its potential?". After the
recent discovery of the Higgs boson at CERN \cite{experiments},
which showed that elementary scalars exist in nature, the most natural and
simplest thing to assume is that inflation is driven
by the Higgs boson (in the standard model or in some extension of
it). Unfortunately it is well known that
the quartic potential, which is the simplest Higgs potential provided by
particle physics in renormalizable theories, has been excluded
by current data \cite{kallosh} since it predicts too many gravity
waves. Although the presence of a non-minimal coupling can make the
quartic potential viable \cite{shafi},
warm inflation provides another solution that is simpler and at the same time,
as we have already mentioned, avoids the discussion about
reheating. If we look at the expressions for the observables in the framework
of warm inflation, we see that the key ingredient that
can in principle reduce the tensor-to-scalar ratio, and bring the predictions
of the model inside the region allowed by observational
data, is the suppression factors $(T/H)$ and $R^{5/2}$. And this is exactly
what happens indeed as we will show in the discussion to follow.

Warm inflation with a quartic potential for the inflaton has also been studied in \cite{BasteroGil:2011xd,papers}.
However there are some differences, as in these works the authors have used
another expression for
the dissipative coefficient,
they have not derived the allowed range for the parameters of the model they
studied, and finally in our work
we have used the most recent data available today.

\subsection{The weak dissipative regime}

Considering our model evolves in agreement with the weak
dissipative regime, where $R\ll1$, and that under the slow-roll approximation the Friedmann
and the Klein-Gordon equations take the standard form, the temperature of the
radiation field assuming an inflaton potential of the form $V(\phi)=(1/4)
\lambda \phi^4$ and an inflaton decay rate $\Gamma=a T$, becomes
\begin{equation}
T\simeq\left(\frac{a V_{,\phi}^2}{36 C_{\gamma} H^3}\right)^{1/3}\label{TW},
\end{equation}
and the Hubble
parameter is given by
\begin{equation}
H\simeq\left(\frac{V}{3M_p^2}\right)^{1/2}\label{HW}.
\end{equation}
In this way, for the weak regime, the slow-roll parameters become
\begin{equation}
\epsilon=\frac{8 M^2_p}{\phi^2},\,\,\,\eta=\frac{12 M^2_p}{\phi^2},\,\,\,\beta=0,\,\,\,\sigma=\frac{4 M^2_p}{\phi^2}.\label{sr}
\end{equation}
It is easy to see that the end of inflation is determined by the condition $\eta = 1$, where the scalar field takes the value $\phi_{\textup{end}}=2\sqrt{3}M_p$.

By the other hand, the number of $e$-folds is given
by the standard formula
\begin{equation}
N=\int_{t_{*}}^{t_{\textup{end}}}\,H\,dt\simeq\frac{1}{M_p^2}\int_{\phi_{\textup{end}}}^{\phi_{*}}\,\frac{V}{V_{\phi}}\,d\phi\simeq \frac{1}{4}\left(\frac{\phi_{*}}{M_p}\right)^2,\label{phiN}
\end{equation}
where we have assumed that $\phi_{*}\gg\phi_{\textup{end}}$.

\begin{figure}[th]
{\hspace{-3
cm}\includegraphics[width=2.8in,angle=0,clip=true]{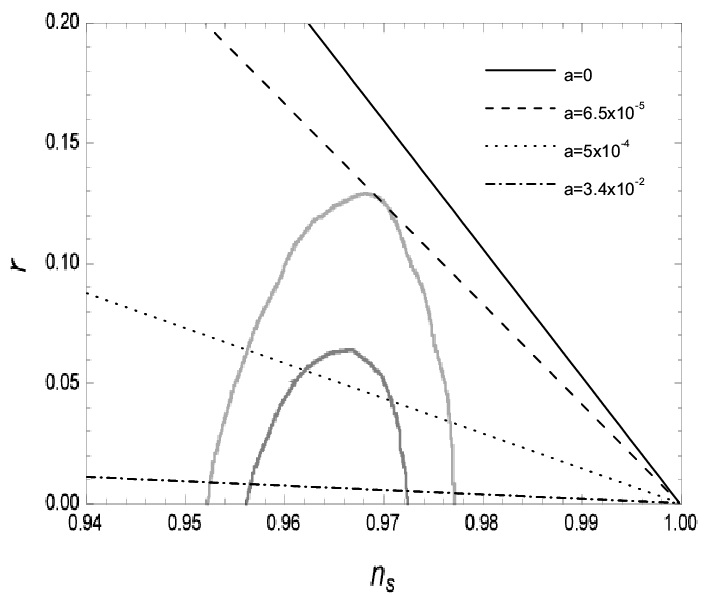}}
{\caption{Plot of the tensor-to-scalar ratio $r$ versus the scalar spectral index $n_s$ in the weak
dissipative regime, for the quartic potential and an inflaton ratio decay $\Gamma=a T$. Here, we have considered the
two-dimensional marginalized joint confidence contours for $(n_s
,r)$, at the $68\%$ and $95\%$ CL, from the latest Planck data \cite{Ade:2015lrj}. In this plot we
have used four different values of the parameter $a$, where the value $a=0$ corresponds to standard cold inflation.
\label{wk1}}}
\end{figure}

In the following, we will study  the scalar and tensor
perturbations. In the weak dissipative regime, the amplitude of the power
spectrum (\ref{PR}) becomes
\begin{equation}
{\cal{P}_{\cal{R}}}^{1/2}\simeq \left(\frac{H}{2 \pi}\right) \left(\frac{3H^2}{V_{,\phi}}\right) \left(\frac{T}{H}\right)^{1/2} .
\end{equation}
By using Eqs.(\ref{TW}), (\ref{HW}), and (\ref{phiN}), it may we written in terms
in the number of $e$-folds as
\begin{equation}
{\cal{P}_{\cal{R}}}^{1/2}\simeq\left(\frac{\lambda \sqrt{a} N^3}{6\sqrt{70}\pi^3}\right)^{1/3}.\label{PNW}
\end{equation}
The power spectrum constraint
$P_R^{1/2} \sim 10^{-5}$ \cite{Ade:2013uln,Ade:2015lrj} determines the dimensionless coupling $\lambda$ in terms of $a$ and $N$, while the
scalar spectral index (\ref{nsw}) turns out to be
\begin{equation}
n_s\simeq 1-2 \epsilon+2 \sigma,
\end{equation}
which may be expressed in terms of the number of the $e$-folds, obtaining
\begin{equation}
n_s=1-\frac{1}{N},\label{nsNw}
\end{equation}
while the tensor-to-scalar ratio (\ref{rwi}) becomes
\begin{equation}
r \simeq \left(\frac{H}{T}\right) 16 \epsilon,\label{rnw}
\end{equation}
so eventually we can obtain
$r$ as a function of $n_s$. Using Eqs.(\ref{PNW}), (\ref{nsNw}), and (\ref{rnw}),
the relation $r(n_s)$ is given by
\begin{equation}
r=\frac{4\sqrt{14}}{625 \sqrt{5}\,a^{1/2}}(1-n_s).
\end{equation}
In figure \ref{wk1}, the relation $r(n_s)$ is shown for several values of $a$.
In the same plot we also show the curve for standard inflation ($a=0$) as well
as the contours allowed by the Planck latest data. When $a$ decreases the
curve is shifted upwards and finally lies outside the allowed contours.
This induces a lower bound on $a$. On the other hand, when $a$ increases the
curve
is shifted downwards, but $R$ also increases and eventually the condition for
being
in the weak dissipative regime is violated. This induces an upper bound on $a$,
which is found to be $6.5\times10^{-5}<a<3.4\times 10^{-2}$.
This implies that the Eq.(\ref{PNW}), evaluated when the cosmological scales
cross the Hubble horizon during inflation at $60$ $e$-folds, gives us
the constraint on $\lambda$ determined by $ 10^{-15}<\lambda<10^{-13}$.
It is interesting to note that this result is in agreement with the value obtained for $\lambda$ in the standard cold inflation
using the COBE normalization \cite{Liddle:2000cg}, given by $\lambda\sim 10^{-14}$.

\subsection{The strong dissipative regime}

Considering our model evolves in agreement with the strong
dissipative regime, where $R\gg1$, under the slow-roll approximation, the temperature of the
radiation field becomes
\begin{equation}
T\simeq\left(\frac{ V_{,\phi}^2}{4 C_{\gamma}\,a H}\right)^{1/5}\label{TWs},
\end{equation}
and the Hubble
parameter is given by Eq.(\ref{HW})
In this way, for the strong regime, the slow-roll parameters become
\begin{equation}
\epsilon=\frac{8 M^2_p}{\phi^2},\,\,\,\eta=\frac{12 M^2_p}{\phi^2},\,\,\,\beta\frac{15 M^2_p}{5\phi^2},\,\,\,\sigma=\frac{4 M^2_p}{\phi^2}.\label{srs}
\end{equation}
For the strong regime, inflation ends when one of these slow-roll parameters becomes the order of $R$. In this case, the end of inflation is determined by the condition $\eta =R$, where the inflaton takes the value $\phi_{\textup{end}}=\frac{(6^{7}35)^{1/4}}{a}\lambda^{1/4}M_p$.

By the other hand, the number of $e$-folds is given
by
\begin{equation}
N=\int_{t_{*}}^{t_{\textup{end}}}\,H\,dt\simeq\frac{1}{M_p^2}\int_{\phi_{\textup{end}}}^{\phi_{*}}\,\frac{V}{V_{\phi}}R\,d\phi\simeq \frac{1}{8}
\left(\frac{a\,5^4}{7\,\lambda 6^2 }\right)^{1/5}\left(\frac{\phi_{*}}{M_p}\right)^{4/5},\label{phiNs}
\end{equation}
where we have assumed that $\phi_{*}\gg\phi_{\textup{end}}$.

Now, the amplitude of the power
spectrum (\ref{PR}) becomes
\begin{equation}
{\cal{P}_{\cal{R}}}^{1/2}\simeq \left(\frac{H}{2\pi}\right) \left(\frac{3H^2}{V_{\phi}}\right)\left(\frac{T}{H}\right)^{1/2}R^{5/4},\label{PRs}
\end{equation}
Similarly to weak regime, the amplitude of the power spectrum may we written in terms
of the number of $e$-folds. Using Eqs.(\ref{HW}), (\ref{TWs}), and (\ref{phiNs}), we have that
\begin{equation}
{\cal{P}_{\cal{R}}}^{1/2}\simeq\left[\frac{4 N^3 \lambda}{125 \pi^{8/3}}\left(\frac{2}{315}\right)^{1/3}\right]^{3/8}.\label{PNs}
\end{equation}
In this case, the power spectrum does not depend on $a$ and then the constraint $P_R^{1/2} \sim 10^{-5}$ \cite{Ade:2013uln,Ade:2015lrj} determines the
inflaton self-interaction coupling $\lambda$.

For this regime, the scalar spectral index (\ref{nsw}) turns out to be
\begin{equation}
n_s\simeq 1+\frac{1}{7R}(-3\eta +18\sigma-9\epsilon),\label{nss}
\end{equation}
which expressed in terms of the number of the $e$-folds yields
\begin{equation}
n_s=1-\frac{45}{28N}.\label{nsNs}
\end{equation}

Finally, for the tensor-to-scalar ratio (\ref{rwi}) we have that
\begin{equation}
r \simeq \left(\frac{H}{T}\right)\frac{16\,\epsilon}{R^{5/2}},\label{rns}
\end{equation}
which may be expressed as function of $n_s$. Using Eqs.(\ref{PNs}), (\ref{nsNs}), and (\ref{rns}),
the relation $r(n_s)$ is given by
\begin{equation}
r\simeq 8.5\times 10^{-9}\frac{\,\pi^{10/3}}{a^4}(1-n_s).
\end{equation}

\begin{figure}[th]
{\hspace{-3
cm}\includegraphics[width=2.8in,angle=0,clip=true]{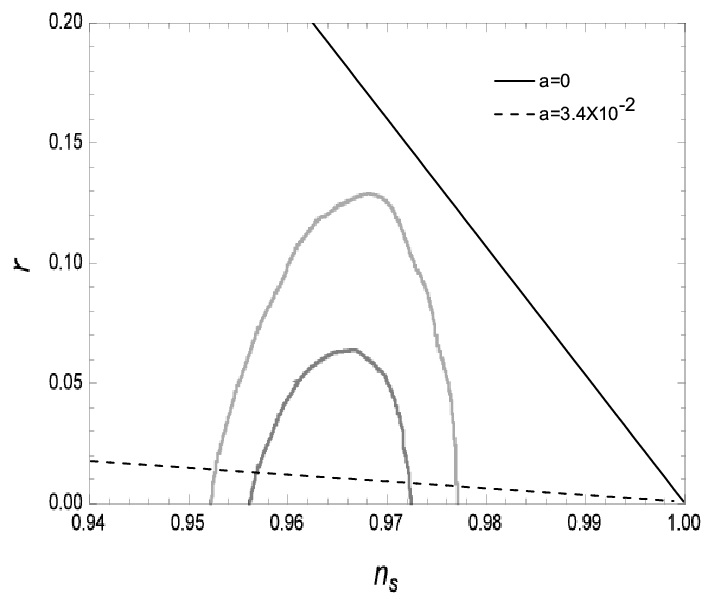}}
{\caption{Plot of the tensor-to-scalar ratio $r$ versus the scalar spectral index $n_s$ in the strong
dissipative regime, for the quartic potential and an inflaton ratio decay $\Gamma=a T$. Here, we have considered the
two-dimensional marginalized joint confidence contours for $(n_s
,r)$, at the $68\%$ and $95\%$ CL, from the latest Planck data \cite{Ade:2015lrj}. In this plot we
have used two different values of the parameter $a$, where the value $a=0$ corresponds to standard cold inflation.
\label{st2}}}
\end{figure}

In figure \ref{st2}, the relation $r(n_s)$ is shown for two different values of $a$.
In the same plot, as in the weak regime, we also show the curve for standard cold inflation ($a=0$) as well
as the contours allowed by the latest Planck data. When $a$ decreases the
curve is shifted upwards, but $R$ also decreases and eventually the condition for
being in the strong dissipative regime is violated. This induces a lower bound on $a$. By
the other hand, when $a$ increases the curve is shifted downwards, but $R$ also increases and the condition for
being in the strong dissipative regime is always satisfied. This implies that there is
only a lower bound for $a$ found by the requirement of staying in the
strong dissipative regime, and given by $a>3.4\times 10^{-2}$. Finally, the Eq.(\ref{PNs}), evaluated at $60$ $e$-folds, gives us
the constraint on $\lambda$, determined by $\lambda\sim 10^{-15}$. This value is almost
the same order that obtained for $\lambda$ in the standard cold inflation.

\section{Conclusions}\label{conclu}

In the present work we have studied warm inflation with a quartic inflaton
potential $V(\phi)=(1/4) \lambda \phi^4$ and an inflaton decay rate
proportional to the temperature, namely $\Gamma=a T$. Warm inflation
consists an alternative to the standard cold inflation, during which radiation is neglected and which requires
two steps, a slow-roll phase followed by a reheating phase, about which very little is known. On the contrary,
in warm inflation, which has the attractive feature that avoids reheating, radiation is also taken into account and it is coupled
to the inflaton leading to testable predictions different than the predictions
of standard inflation even in the weak dissipative regime. The model we have
considered is characterized by two parameters, namely the dimensionless
couplings $a$ and $\lambda$. We have used the latest Planck data to constrain
the parameters of the model, and the results we have obtained are shown in the
figures \ref{wk1} and \ref{st2} for the case of weak and strong dissipative regime
respectively. In the weak regime first, where $\Gamma \ll 3H$, the background equations
look the same as in standard inflation, however the tensor-to-scalar ratio
is suppressed by the factor $T/H$, which must always be larger than one in
warm inflation. The power spectrum constraint determines $\lambda$ in terms of
$a$, and then the tensor-to-scalar ratio as a function of the scalar index
$n_s$ changes according to $a$ as follows: As $a$ increases the theoretical curve
is shifted downwards, and on the other hand as $a$ decreases the theoretical
curve is shifted upwards. We have obtained both an upper and a lower bound on
$a$, since when $a$ becomes too low the theoretical curve lies outside the
contours allowed by data, and when $a$ becomes too large the condition for
the weak dissipative regime is not satisfied. In figure \ref{wk1} we show the
contours allowed by the data together with four theoretical curves, namely
one for the standard inflation and for three
different values of the coupling $a$ in warm inflation, the minimum value, the maximum value and
one intermediate value. In the strong dissipative regime, where $\Gamma \gg 3H$, the
power spectrum does not depend on $a$ and so the constraint determines the
inflaton self-interaction coupling $\lambda$. In the figure \ref{st2}, the $r-n_s$ plot is shown
and there is only a lower bound for $a$, obtained by the requirement of staying in the
strong dissipative regime. In this regime the tensor-to-scalar ratio is
suppressed by the factor $T/H$ as in the weak regime, but also by the factor
$R^{5/2}$.
That is why in the strong regime the model always predicts a very low $r$.
By the other hand, we observe that the constraints found on the coupling $\lambda$, in
both dissipative regimes, are in agreement with
the value obtained in standard cold inflation using the COBE normalization.
In this way, we conclude that warm inflation can rescue the quartic potential that in
standard inflation is ruled out by the data.


\begin{acknowledgments}
The authors would like to thank G. Barenboim for helping us with the figures. G.P. was supported by Comisi\'on Nacional
de Ciencias y Tecnolog\'ia of Chile through
Anillo project ACT1122. N.V. was supported by Comisi\'on Nacional
de Ciencias y Tecnolog\'ia of Chile through FONDECYT Grant N$^{0}$
3150490. Finally, we wish to thank the anonymous referee for her/his valuable comments,
that have helped us to improve the presentation of our manuscript.
\end{acknowledgments}


\end{document}